\documentclass[pdflatex,iicol,sn-mathphys-num]{sn-jnl}  

\usepackage{graphicx}
\usepackage{multirow}
\usepackage{amsmath,amssymb,amsfonts}
\usepackage{amsthm}
\usepackage{mathrsfs}
\usepackage[title]{appendix}
\usepackage{xcolor}
\usepackage{textcomp}
\usepackage{manyfoot}
\usepackage{booktabs}
\usepackage{algorithm}
\usepackage{algorithmicx}
\usepackage{algpseudocode}
\usepackage{listings}
\usepackage{subfigure}

\begin{document}

\title[mode=title]{Holographic interference surface: A proof of concept based on the principle of interferometry}

\author*[1]{\fnm{Haifan} \sur{Yin}}\email{yin@hust.edu.cn}
\author[1]{\fnm{Jindiao} \sur{Huang}}\email{jindiaohuang@hust.edu.cn}
\author[1]{\fnm{Ruikun} \sur{Zhang}}\email{rkzhang@hust.edu.cn}
\author[1]{\fnm{Jiwang} \sur{Wu}}\email{wujiwang@hust.edu.cn}
\author[1]{\fnm{Li} \sur{Tan}}\email{ltan@hust.edu.cn}

\affil*[1]{\orgdiv{School of Electronic Information and Communications}, \orgname{Huazhong University of Science and Technology}, \orgaddress{\city{Wuhan}, \postcode{430074}, \country{China}}}

\abstract{
Revolutionizing communication architectures to achieve a balance between enhanced performance and improved efficiency is becoming increasingly critical for wireless communications as the era of ultra-large-scale arrays approaches. In traditional communication architectures, radio frequency (RF) signals are typically converted to baseband for subsequent processing through operations such as filtering, analog-to-digital conversion and down-conversion, all of which depend on expensive and power-intensive RF chains. The increased hardware complexity and escalated power consumption resulting from this dependency significantly limit the practical deployment of ultra-large-scale arrays. To address these limitations, we propose a holographic communication system based on the principle of interferometry, designated as holographic interference surfaces (HIS). Utilizing the interference effect of electromagnetic waves, HIS estimates the channel state information (CSI) by dealing solely with power information, which enables the replacement of RF chains with power sensors and completes the signal processing in radio frequency. As proof-of-concept demonstrations, we implemented a prototype system based on principles of holographic interference. Experimental results align well with theoretical predictions, confirming the practical viability and effectiveness of the proposed HIS. This work provides a new paradigm for building a more cost-effective wireless communication architecture.
}

\maketitle

\section{Introduction} \label{secIntro}
Along with the arrival of the sixth generation (6G) communication era, the demand for the intelligent connection of everything and the seamless integration of communication is becoming increasingly urgent, requiring Tbps-scale data rates, Kbps/Hz-scale spectral efficiency and $\mu$s-level latency \cite{Chen:23JSAC, Chettri:20ITJ, Saad:20Network, You:23WC}. Considering the superior performance and tremendous potential of massive multiple-input multiple-output (MIMO) technology demonstrated in the fifth generation (5G) communications \cite{Marzetta:10TWC, Larsson:14CM, Swindlehurst:14CM}, deploying extremely large-scale arrays to support the construction of 6G is currently a mainstream solution \cite{Wang:23CST, Dang:20NE}. However, the issues associated with the growing number of antennas, such as the expanded array size, increased hardware complexity and escalated power consumption, dramatically hinder the deployment of extremely large-scale arrays in practical scenarios \cite{Bjornson:17FTSP, Liu:24JSAC}.

To mitigate the issues, researchers have conducted extensive explorations from the aspects of array structures and communication architectures \cite{ Ye:24CL, Huang:24WCL, You:23TWC, Xu:24TWC, An:23JSAC}. Holographic multiple-input multiple-output (HMIMO), as an extension to massive MIMO, is one of the immensely promising technologies \cite{Gong:24CST, An:23CL, Dardari:20JSAC}. The core concept of HMIMO is enabling dense or quasi-continuous antenna deployment in limited space by compressing the spacing between each antenna to within half a wavelength \cite{Pizzo:20JSAC, Huang:20WC, Wan:21TCOM, Gong:24WC}. The above hardware implementation is made feasible through metasurfaces, which manifest as nearly spatially continuous apertures and provide two key characteristics for manipulating electromagnetic (EM) waves. By utilizing the mutual coupling effects caused by the sub-wavelength distances between two adjacent antenna elements, HMIMO surfaces are capable of forming pencil-like superdirective beams \cite{Damico:23TWC, Han:24OJCS}. Besides, the continuous apertures enable HMIMO to record the continuous phase changes of the wavefront, thus providing support for the complex multi-dimensional manipulation of EM waves \cite{Gong:24CST}. All of these features enable HMIMO to fully exploit the capacity potential of MIMO systems in a limited space, thereby improving system performance to meet 6G requirements \cite{An:23CL_P3}. However, the source of power consumption and hardware complexity in massive MIMO systems lies in the costly and power-hungry radio frequency (RF) chains, implying that HMIMO is fundamentally incapable of addressing the issues of increased hardware complexity and escalated power consumption in extremely large arrays \cite{Huang:24TWC, Zhu:23TIT}.

Under traditional signal processing architectures, RF signals are typically converted to baseband for subsequent processing, indicating that communication systems remain reliant on RF chains as long as the underlying logic of signal processing is preserved. Therefore, a new wireless signal transceiver principle is needed to construct an architecture that can perform signal processing in radio frequency, significantly reducing the reliance on RF chains. In order to better explain this new signal transceiving criterion, we now introduce a technology for high-fidelity recording and reconstruction of optical wavefields, known as holography \cite{Hariharan:96, Schnars:15, Fratz:21LAM}. Introduced by Gabor in 1948 \cite{Gabor:48Nat}, holography uses the interference of two coherent optical waves to record complete wave information while still employing conventional optical sensors. It overcomes the limitation of traditional photography, which cannot capture the phase of the optical wavefield. Considering the similarity in the physical nature of light and EM waves, the holographic interference principle has the potential to be extended to wireless communications. The inspiration of holographic interference principles for wireless communication lies in their ability to receive and process signals by dealing solely with power information \cite{Collier:13}. It indicates that complex operations, such as filtering, down-conversion, and analog-to-digital conversion (ADC) in RF chains can be replaced by power sensors like envelope detectors. This enables operations to be shifted from baseband to radio frequency for EM-domain signal processing, resulting in a significantly simplified hardware design and reduced power consumption.

In our previous works \cite{Huang:24TWC, Huang:25arxiv}, we extended the principle of holographic interference to wireless communications and conceived a fundamentally different multi-antenna system, termed holographic interference surface (HIS), which exploits interference effects of RF signals. Leveraging the 3GPP cluster delay line channel model \cite{3GPP:TR38901}, we analyzed the signal processing flow enabled by the holographic interference theory, on the basis of which we further proposed a holographic channel sensing architecture for RF hologram-based channel estimation \cite{Huang:24TWC}. The feasibility of holographic channel sensing in wideband scenarios was also examined \cite{Huang:25arxiv}. We constructed a transformation relation between holograms and sinusoidal signals from a perspective of geometry and derived the Cram\'er-Rao lower bound (CRLB) for the wideband channel estimation problem.

\begin{figure}[h!]
    \centering
    \includegraphics[width=\columnwidth]{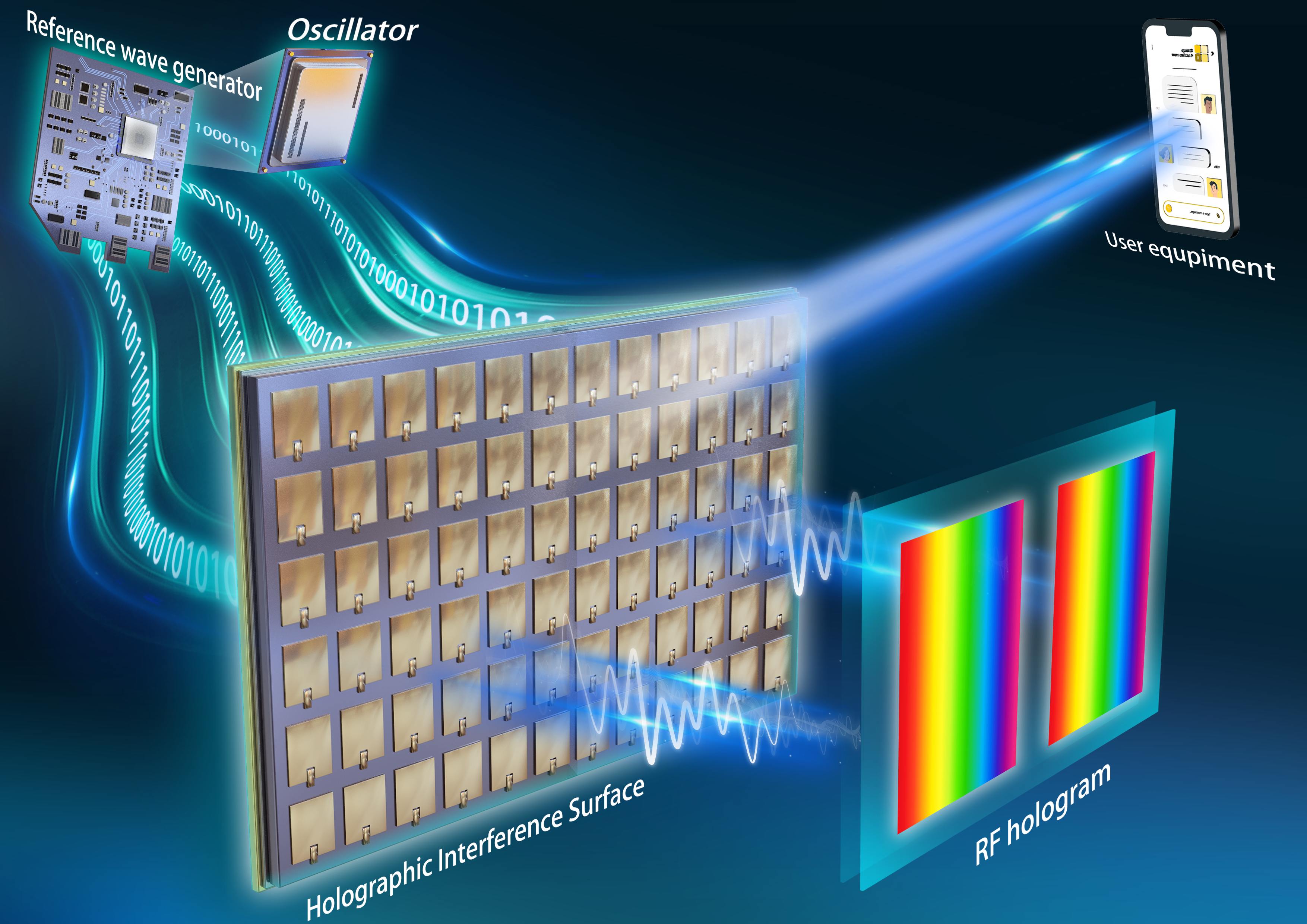}
    \caption{\textbf{Schematic diagram of the holographic interference surface.} The UE-transmitted signal and the surface-generated reference signal are interfered on the unit through combiners. Subsequently, the envelope detector captures the power of the superimposed signal to generate the radio frequency hologram, which is then transmitted to the control board for channel estimation.}
    \label{figSchDiag}
\end{figure}

In this article, we design the hardware structure of HIS and experimentally demonstrate this holographic interference principle-based communication system. The HIS, composed of multiple metamaterial radiation units, detects or estimates wireless signals based on the interferometric superposition of the object wave transmitted by the user equipment (UE) and the reference wave generated by the surface. The three-dimensional electromagnetic space is then perceived and reconstructed efficiently by HIS based on the holograms. With its excellent RF spatial awareness and spatial wave field synthesis capability, the HIS can achieve precise sensing, accurate regulation, and intelligent optimization of the electromagnetic environment, which is likely to provide a revolutionary communication architecture for building a more cost-effective system. For experimental validation, we implemented a prototype system based on holographic interference principles. A 32-unit array was designed and fabricated, with each of its units equipped with combiners and power detectors to enable holographic channel sensing based on the power distribution of RF signals. The experimental results are in strong agreement with theoretical simulations, thereby confirming the practical viability and effectiveness of the proposed HIS.

\section{Results} \label{secRes}
\subsection{Principle of holographic communication}
Holography, originally developed within the domain of optics, is an imaging methodology that exploits the interference and diffraction phenomena of coherent light to achieve high-fidelity three-dimensional recording and reconstruction. The optical holographic process is formally decomposed into two sequential phases: (i) interferometric encoding and (ii) diffractive decoding. During interferometric encoding, the complex amplitude distribution of the object wave is coherently superposed with a known reference wave, yielding an intensity pattern termed the hologram that encodes both amplitude and phase information of the original field. The hologram, subsequently recorded on a suitable photosensitive medium, constitutes a complete spatial–frequency representation of the object wave. In the diffractive decoding stage, the hologram is illuminated by a reconstruction wave (typically a replica of the original reference wave), resulting in the diffraction-mediated regeneration of the object wavefront and thereby reconstructing the three-dimensional image.

Owing to the formal analogy between optical and RF electromagnetic fields, the principles underlying optical holography can be generalized to wireless communication systems. Specifically, the electromagnetic signal can be spatially encoded as an intensity distribution over a receiving aperture, thereby enabling information retrieval and retransmission through holographic techniques. This conceptual extension suggests a paradigm in which the holographic framework serves as a unified platform for both imaging and communication functionalities. In order to keep the consistency of terms in optical and communication holography, we denote the signal transmitted by the UE as the object wave, and the sinusoidal signal generated by HIS as the reference wave.

In the holographic communication architecture, electromagnetic-wave reception is characterized by the interferometric superposition of two coherent beams. Their interaction yields a spatial interference pattern, called hologram, which simultaneously encodes the amplitude and the relative phase of the object wave. The HIS leverages this phenomenon: it forms a hologram through the coherent superposition of an object wave and a reference wave, and subsequently reconstructs the object-wave information from the hologram and the reference wave to accomplish signal reception. For an HIS comprising $N_v \times N_h$ radiating unit, the generated reference wave and the object wave incident upon the $(p,s)$-th unit are expressed as:
\begin{align}
    E_{r}(\mathbf{r}_{p,s},t) 
        & = A_{r} \cdot e^{j (\mathbf{k}_{\text{r}} \cdot \mathbf{r}_{p,s} + \omega_rt)}, \\
    E_{o}(\mathbf{r}_{p,s},t) 
        & = A_{o} \cdot e^{j (\mathbf{k}_{o} \cdot \mathbf{r}_{p,s} + \omega_ot)},
\end{align}
where $A_{r}$ and $A_{o}$ denote the scalar amplitudes of the locally generated reference wave and the incident object wave, respectively. $\mathbf{r}_{p,s}$ represents the spatial position vector of the $(p, s)$-th unit. $\mathbf{k}_{o}$ is the free-space propagation vector associated with the object wave, while $\mathbf{k}_{r}$ corresponds to that of the reference wave. The coherent superposition of these two waves yields the interference field, whose instantaneous power distribution, namely the hologram, is given by:
\begin{align} \label{eqHGM}
    &E_I(\mathbf{r}_{p,s},t) \nonumber \\
    &= |E_o(\mathbf{r}_{p,s},t) + E_r(\mathbf{r}_{p,s},t)|^2 \nonumber \\
    &= |E_o(\mathbf{r}_{p,s},t)|^2 + |E_r(\mathbf{r}_{p,s},t)|^2 \nonumber \\
    &+ \! E_o(\!\mathbf{r}_{p,s},t) \! E_r^*(\!\mathbf{r}_{p,s},t) \! + \! E_o^*(\!\mathbf{r}_{p,s},t)\! E_r(\!\mathbf{r}_{p,s},t).
\end{align}

Eq. (\ref{eqHGM}) reveals that the hologram arising from interferometric superposition comprises a direct-current (DC) term proportional to the sum of the instantaneous powers of the two waves and a pair of complex-conjugate terms that encode the phase difference between the two waves. Because the reference wave is locally generated by the HIS, its amplitude and phase are precisely known. Accordingly, the complex envelope of the object wave can be recovered solely from the hologram and the reference-wave parameters, thereby completing signal reception.

As indicated by Eq. (\ref{eqHGM}), the hologram comprises a DC term together with a conjugate pair. If the hologram is illuminated by the original reference wave and directly applied for beamforming without further processing, then the resulting field distribution on the HIS plane is given by
\begin{align} \label{eqREC}
    E_{c} 
        =& E_{r} \cdot E_I \nonumber \\
        =& (A_{r}^2 \!+\! A_{o}^2)E_{r} \!+\! A_{r}^2E_{o} \!+\! A_{r}^2e^{j2 \mathbf{k}_{r} \cdot \mathbf{r}_{p,s}}E_{o}^*.
\end{align}

Eq. (\ref{eqREC}) indicates that the reconstructed wave field comprises three constituents: an amplitude-modulated replica of the reference wave, the original object wave, and a phase-modulated conjugate of the object wave. 
In this configuration, the antenna array employs a local oscillator to produce the reference wave. This wave is distributed via power dividers and a microstrip network to ensure identical amplitude and phase excitation at each unit. 
Consequently, the second term in Eq. (\ref{eqREC}) reduces to an amplitude-modulated wave that propagates along the trajectory of the primary object wave $E_o$. The third term retains the conjugate waveform of $E_o$ multiplied by an additional phase factor. Although this conjugate component exhibits beam characteristics analogous to those of the original object wave, its propagation direction is governed solely by the phase factor. Therefore, the unprocessed hologram fails to produce the intended beamforming performance. Instead, the DC and conjugate terms act as coherent interference sources that degrade the desired wave field.

Although holographic communication and optical holography are founded on identical physical principles, their signal processing demands differ in several critical respects. These differences primarily stem from the disparate application environments, requirements, and media \cite{Huang:24TWC, Huang:25arxiv}. In contrast to the stable and homogeneous spectral composition in optical photography, the widespread presence of broadband noise and interference in communication environments leads to an extremely complex RF spectral composition, further exacerbated by multipath effects in complex spaces and Doppler effects in mobile environments. From a requirements perspective, optical holography has limited demand for imaging in high-speed mobility scenarios, which are relatively common in 6G communication \cite{Yin:20JSAC}. This places higher demands on communication holography in terms of real-time performance and stability to address the frequent and intense changes in the EM environment. Finally, by leveraging the nanometer-scale wavelengths of light waves, optical arrays typically contain a massive number of sensing units, enabling the acquisition of spatial domain information with ultra-high spatial resolution for optical holography \cite{Kreiss:05}. However, the spatial information available for communication holography is significantly less than that of optical holography due to limitations in unit size and EM wavelength, necessitating greater reliance on time-frequency domain samples to enhance signal processing efficiency and accuracy.

Optical hologram interference-suppression strategies are conventionally categorized into five principal classes: phase-shifting interferometry (PSI) \cite{Yamaguchi:97OL, Lai:91JOSAA, Deck:03AO}, off-axis recording geometries \cite{Leith:63JOSA, Cuche:00AO}, linear spatial filtering \cite{Onural:87OE}, iterative phase retrieval \cite{Latychevskaia:07PRL}, and complex-wavefront filtering in the reconstruction plane \cite{Pedrini:98AO}. Among these, only PSI operates without dependence on high-resolution spatial sampling or perfect channel state information (CSI), both of which are unavailable in communication scenarios. 
By acquiring a small number of phase-shifted hologram samples, PSI simultaneously eliminates the DC term and the conjugate term, thereby satisfying the real-time, low-resolution, and CSI-deficient constraints of holographic channel estimation. 
Accordingly, the interference-suppression framework described below is derived based on PSI.

Assume that the carrier frequency of each UE is identical to that of the reference wave, and the sampling interval is an integer multiple of the reference wave period. Without considering the movement of users, the object wave emitted by the UE is static in each sampling window. Prior to each acquisition, the reference wave is phase-shifted by $\pi/2$, yielding three phase-diverse holograms after three consecutive samples:
\begin{subequations}
    \begin{align}
        I(0)  
            &= |E_{o} + E_{r}|^2 \nonumber \\
            &= |E_{o}|^2 + |E_{r}|^2 + E_{o}E_{r}^* + E_{o}^*E_{r},\\
        I(\frac{\pi}{2})
            &= |E_{o} + jE_{r}|^2 \nonumber \\
            &= |E_{o}|^2 + |E_{r}|^2 - jE_{o}E_{r}^* + jE_{o}^* E_{r},\\
        I(\pi)
            &= |E_{o} - E_{r}|^2 \nonumber \\
            &= |E_{o}|^2 + |E_{r}|^2 - E_{o}E_{r}^* - E_{o}^*E_{r}.
    \end{align}
\end{subequations}
According to the above equations, the conjugate term is eliminated by forming a complex sum of the two acquired holograms:
\begin{subequations}
    \begin{align}
        I(0) \!+\! j\! \cdot\! I(\frac{\pi}{2}) 
            &= (1 \!+\! j)(A_r^2 \!+\! A_o^2) \!+\! 2 E_{o}E_{r}^*,\\
        I(\frac{\pi}{2}) \!+\! j \!\cdot\! I(\pi) 
            &= (1 \!+\! j)(A_r^2 \!+\! A_o^2) \!-\! 2 E_{o}E_{r}^*.
    \end{align} 
\end{subequations}
Therefore, the object wave is recovered with three designed holograms as:
\begin{align} \label{eqPSIS}
    \hat{E}_{o} 
        = \frac{1\!-\!j}{4E_{r}^*} \left\{I(0) \!-\! I(\frac{\pi}{2}) \!+\! j\left[I(\frac{\pi}{2}) \!-\! I(\pi)\right]\right\}.
\end{align}
Under the assumption of a static channel, the object wave can be reconstructed precisely from a finite set of phase-shifted holograms. 

\begin{figure*}[t!]
    \centering
    \includegraphics[width=0.7\textwidth]{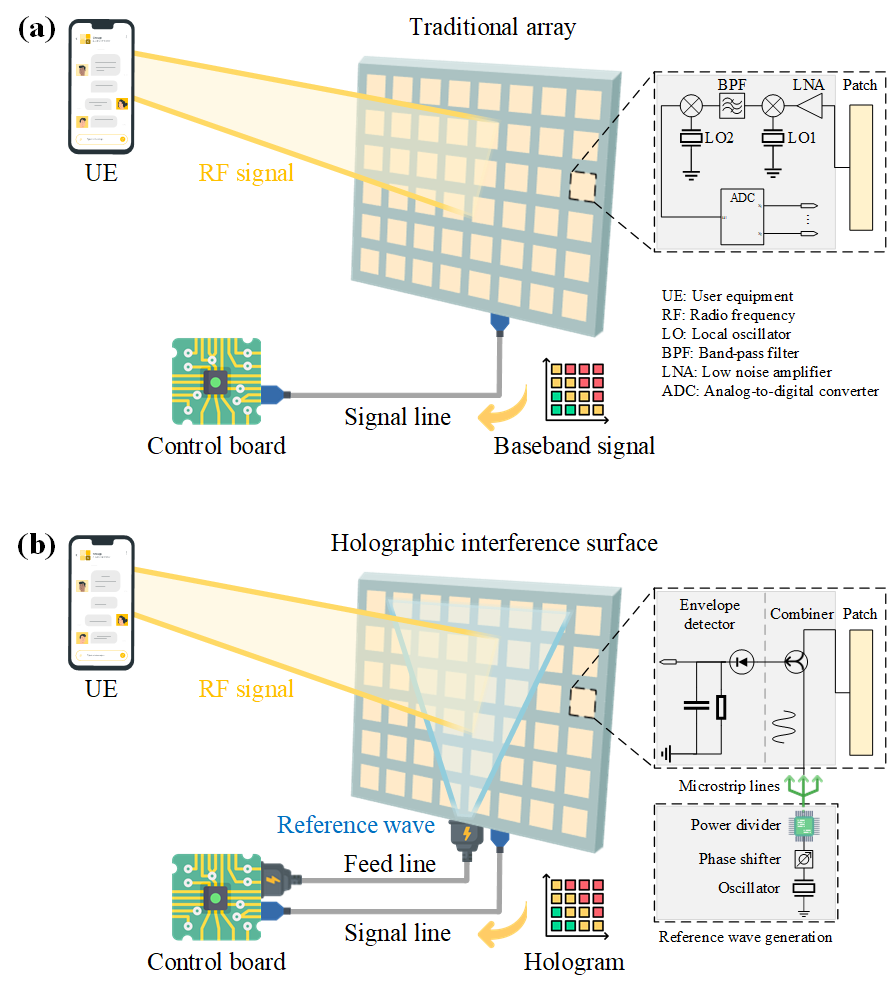}
    \caption{\textbf{Schematic diagrams and array structures of conventional MIMO receiver and the proposed HIS.} \textbf{(a)} Signal processing flow and hardware composition of conventional array. \textbf{(b)} RF signal processing structure and hardware design of the proposed HIS.}
    \label{figArchComp}
\end{figure*}

Traditional multiple-antenna architectures estimate the channel by means of dedicated RF chains followed by baseband processing, an approach whose cost and complexity scale unfavorably with the number of antennas. In contrast, the proposed holographic system performs channel estimation directly in the RF domain. Specifically, the channel state information is captured via RF holograms, thereby eliminating the need for per-element RF chains and the associated filtering, analog-to-digital conversion, and down-conversion stages. This strategy offers a fundamentally new and cost-efficient paradigm for realizing ultra-large-scale wireless communication systems.

\subsection{Holographic interference surfaces}
Fig. \ref{figArchComp}a illustrates the schematic framework of a conventional receiver architecture. In traditional communication systems, RF signals are converted into digital baseband signals through a series of operations, including filtering, down-conversion, and analog-to-digital conversion. Since the signals received by each antenna element must be converted to baseband for subsequent processing, each element requires a dedicated RF chain capable of performing these operations. However, RF chains typically consist of high-power components such as amplifiers, filters, and oscillators. Consequently, increasing the number of antennas enhances both the array gain and the overall system power consumption. The authors of \cite{Bjornson:17FTSP} investigated the trade-off between array gain and power consumption, demonstrating that exceeding a certain number of antennas significantly increases hardware complexity and power usage, leading to a sharp decline in energy efficiency. With the rapid advancement of applications such as extended reality, autonomous driving, and smart cities, expanding antenna arrays to meet the communication demands of these scenarios has become a prevailing trend. Therefore, there is an urgent need for a novel communication architecture that can overcome the power consumption limitations imposed by ultra-large-scale antenna arrays.

In order to improve the energy efficiency of the system in channel estimation, we propose a novel system holographic interference-based channel sensing architecture. By migrating signal processing directly to the RF domain, the proposed scheme drastically reduces the required number of RF chains for channel estimation. The operational principle of the proposed HIS is illustrated in Fig. \ref{figArchComp}(b). For channel estimation, the HIS first generates a sinusoidal signal using a voltage-controlled oscillator to serve as a reference. This reference signal is then split into multiple signals of equal amplitude by a power divider and distributed to each unit via microstrip lines. At each unit, the RF signal received by the patch antenna is combined with the reference signal using a combiner. The power of the resulting superimposed signal is detected by an envelope detector as a hologram. These holograms are subsequently transmitted to the control board, where they are analyzed to estimate the CSI.

Comparing Fig. \ref{figArchComp}(a) and Fig. \ref{figArchComp}(b), it can be observed that the proposed architecture replaces the mixer with a combiner and employs envelope detectors for RF signal processing, thereby eliminating a large number of high-power-consumption components such as filters, amplifiers, and oscillators. In summary, the RF chains originally attached to each unit are replaced with power detection devices. The proposed architecture reduces the total number of RF chains required for channel estimation to one, achieving independence between array size and the number of RF chains. This design overcomes the limitations of traditional MIMO systems in terms of power consumption and hardware complexity, which offers strong potential for the effective deployment of ultra-large-scale arrays within the broader framework of green communications. The detailed design and performance of the HIS units and arrays are presented in the following.

\begin{figure*}[t!]
    \centering
    \includegraphics[width=\textwidth]{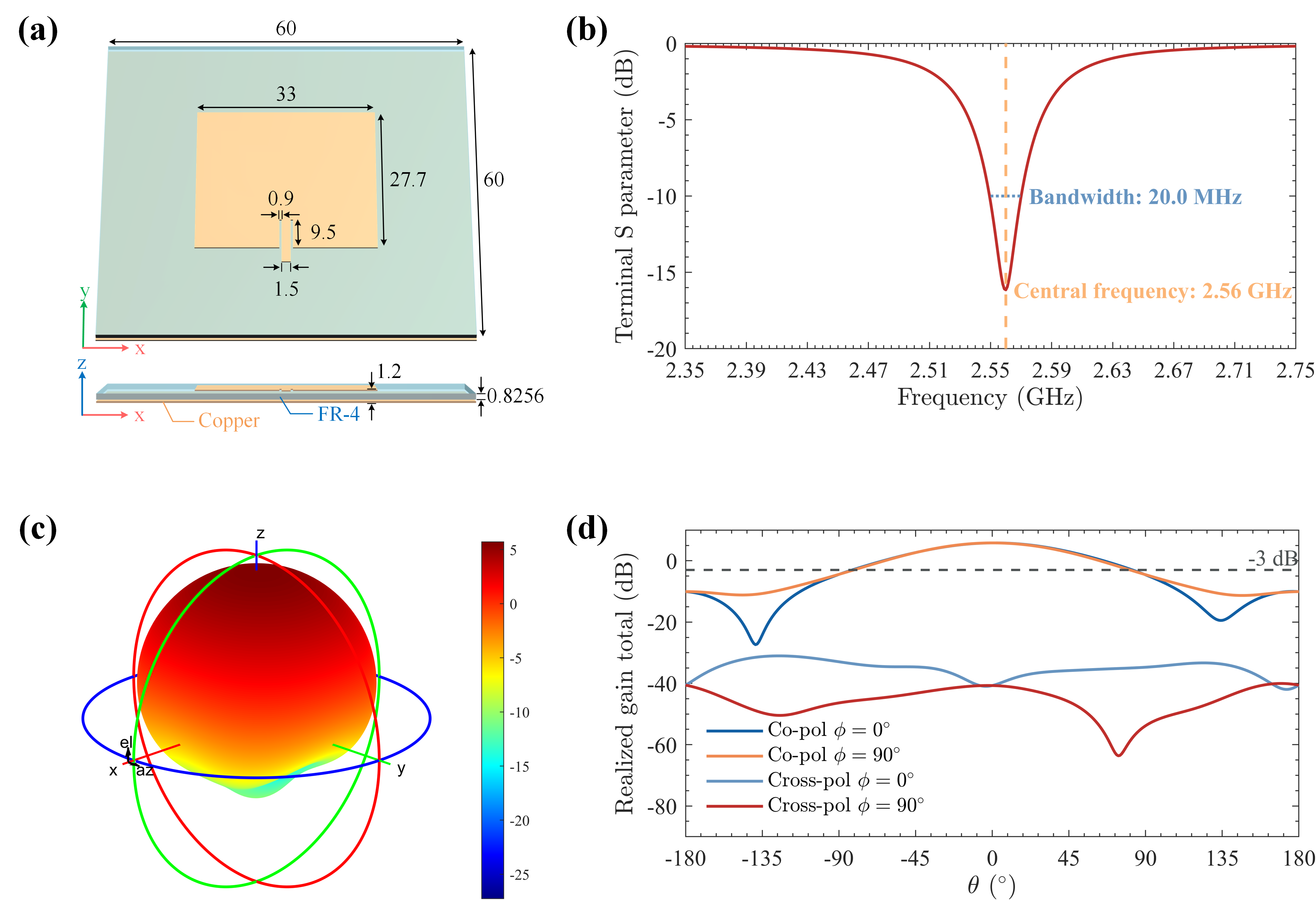}
    \caption{\textbf{Details of the microstrip antenna used in HIS.} \textbf{(a)} Front and cross-section views of the antenna unit. All geometrical parameters are in mm. \textbf{(b)} Simulation of the S-parameter. \textbf{(c)} 3D radiation pattern of the antenna. \textbf{(d)} Co-polarization and cross-polarization characteristics.}
    \label{figHISUnit}
\end{figure*}

The structural design, S-parameters, and unit gain of the antenna employed in the HIS are presented in Fig. \ref{figHISUnit}. The dimensions of the microstrip antenna are designed as 60 mm $\times$ 60 mm $\times$ 1.2 mm. This antenna unit comprises three layers: the middle layer is an FR-4 dielectric substrate with a thickness of 0.8256 mm, and the bottom layer is a copper ground plane. Fig. \ref{figHISUnit}(a) illustrates the detailed structure and specific geometric parameters of the microstrip antenna. The antenna is designed to operate at a center frequency of 2.56 GHz with a bandwidth of 20 MHz. These design specifications are validated by the S11 parameters shown in Fig. \ref{figHISUnit}(b). The radiation patterns of the antenna are depicted in Fig. \ref{figHISUnit}(c) and Fig. \ref{figHISUnit}(d). Additionally, the co-polarization and cross-polarization characteristics are illustrated in Fig. \ref{figHISUnit}(d), from which the horizontal half-power beamwidth, vertical half-power beamwidth, and cross-polarization discrimination are determined to be $161^\circ$, $164^\circ$, and 36.85 dB, respectively.

\begin{figure*}[t!]
    \centering
    \includegraphics[width=\textwidth]{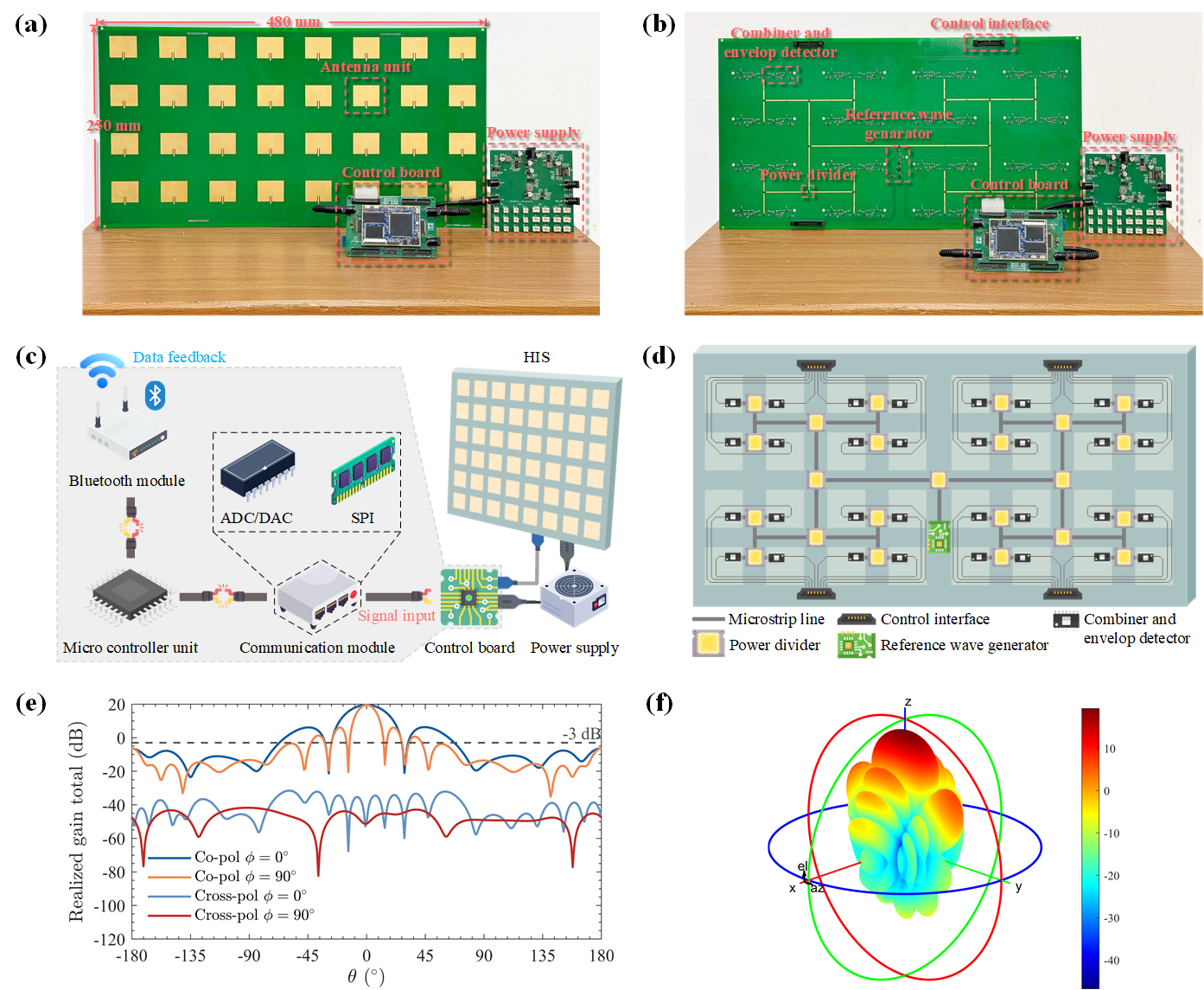}
    \caption{\textbf{The structure and performance of the proposed HIS.} \textbf{(a)} Front view and configuration of the system. \textbf{(b)} Back view and configuration of the system. \textbf{(c)} Composition and control logic of the control board. \textbf{(d)} The back structure and configuration details of HIS. \textbf{(e)} 2D radiation pattern and polarization characteristics of HIS. \textbf{(f)} 3D radiation pattern of HIS.}
    \label{figHISArray}
\end{figure*}

Fig. \ref{figHISArray} illustrates the array structure, control logic, and overall performance of the HIS. As shown in Fig. \ref{figHISArray}(a), the HIS primarily consists of three components: the antenna array, the control board, and the power supply module. The array, with dimensions of 250 mm $\times$ 480 mm, comprises a total of 32 antennas, as depicted in Fig. \ref{figHISUnit}. On the backside of the array, the power detection module, the reference wave generation module, and the control interfaces are integrated. The detailed layout of the rear side of the array is presented in Fig. \ref{figHISArray}(b) and Fig. \ref{figHISArray}(d).

\begin{figure*}[t!]
    \centering
    \includegraphics[width=\textwidth]{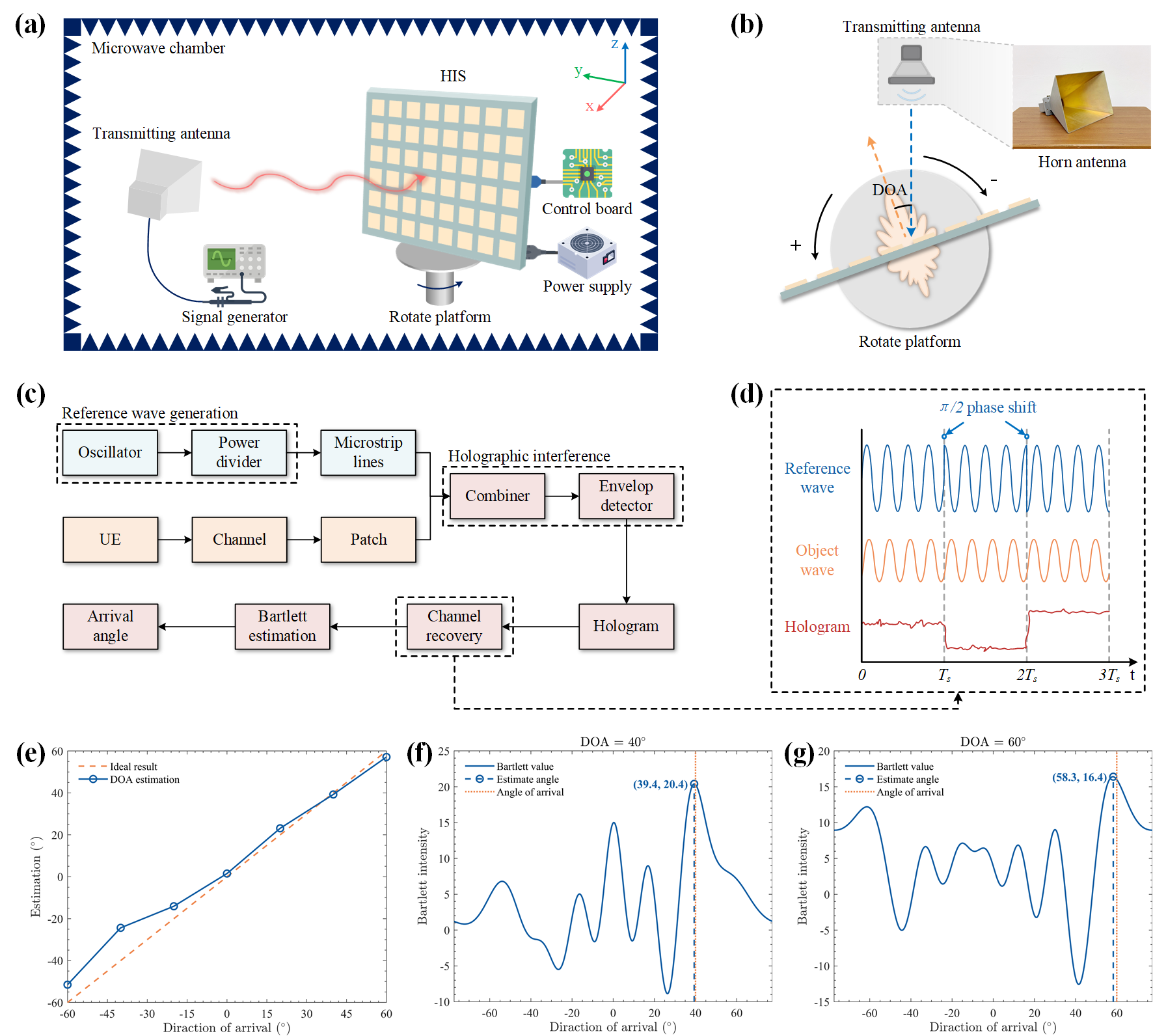}
    \caption{\textbf{Experimental setup and direction of arrival (DOA) estimation results.} \textbf{(a)} Measurement setup in a microwave anechoic chamber. \textbf{(b)} Schematic of the experimental setup to evaluate the channel estimation performance of HIS. \textbf{(c)} Signal processing flow of holographic channel sensing. \textbf{(d)} Schematic of the channel recovery. \textbf{(e)} Results for DOA estimation by using holographic interference principles. \textbf{(f), (g)} Bartlett estimation results at different DOA, such as \textbf{(f)} $\theta = 40^\circ$ and \textbf{(g)} $\theta = 60^\circ$.}
    \label{figExpSet}
\end{figure*}

The reference wave generator is capable of producing a 2.6 GHz RF signal using a voltage-controlled oscillator, a digital attenuator, and a digital phase shifter. The generated reference signal is subsequently divided into multiple channels by a T-shaped power divider network. This power divider splits the reference signal into different channels with equal amplitude and phase. Microstrip lines are employed to deliver the divided signals to antenna units. Behind each unit, combiners and envelope detectors are installed to capture the RF holograms. The envelope detector converts the power of the superimposed signal into an analog voltage, which is then transmitted to the control board through the control interfaces for further processing.

\begin{figure*}[t!]
    \centering
    \includegraphics[width=.75\textwidth]{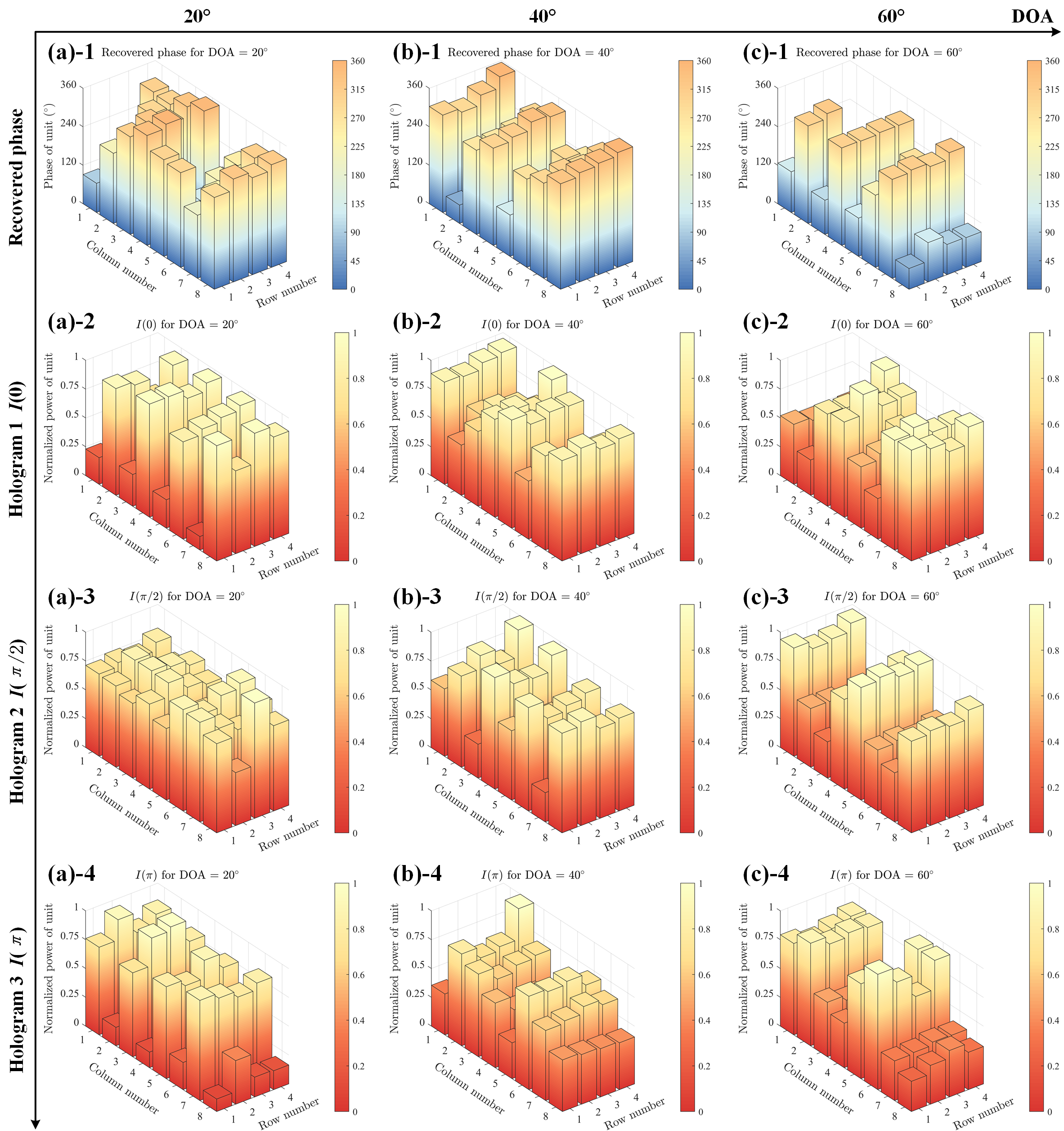}
    \caption{\textbf{Measured results of channel recovery at different direction of arrival (DOA).} \textbf{(a), (b), (c)} Recovered phase of each unit and original holograms of channel recovery for different DOAs.}
    \label{figMeas}
\end{figure*}

The structure and control logic of the control board are illustrated in Fig. \ref{figHISArray}(c). The control board primarily comprises three modules: a microcontroller unit (MCU), a communication module, and a Bluetooth module. The MCU serves as the core computational and control unit, responsible for executing the holographic interference suppression algorithm and the Bartlett estimation method. The communication module integrates independent analog-to-digital converters (ADCs), digital-to-analog converters (DACs), and a serial peripheral interface (SPI). The DAC is utilized to convert the digital control signals from the MCU into analog voltages, which are then used to drive the voltage-controlled oscillator in the reference wave generator for generating a 2.6 GHz sinusoidal signal. To enable external control and data exchange, the control board is also equipped with a Bluetooth module, allowing the HIS to communicate wirelessly with mobile devices or computers.

The 2D and 3D radiation patterns of the array are presented in Fig. \ref{figHISArray}(e)–(f). Based on the co-polarization and cross-polarization characteristics shown in Fig. \ref{figHISArray}(e), the horizontal half-power beamwidth, vertical half-power beamwidth, and cross-polarization discrimination of the array are $26^\circ$, $52^\circ$, and 51.03 dB, respectively. The array exhibits a relatively narrow main lobe beamwidth and a high cross-polarization discrimination level, ensuring effective isolation between co-polarized and cross-polarized components.

\subsection{Experimental results}
To evaluate the channel estimation performance of the HIS, a series of experiments were conducted in a microwave anechoic chamber, as illustrated in Fig. \ref{figExpSet}(a)–(b). A pyramidal horn antenna was fixed at the transmitting station to emulate user equipment. Meanwhile, the HIS was mounted on a rotatable platform, which was used to adjust the angle of arrival of the signal through rotation.

The overall signal processing flow of the experiment is illustrated in Fig. \ref{figExpSet}(c). The signal transmitted by the UE arrives at the patch antenna after propagating through the channel, while the reference signal generated by the HIS is distributed to the units via microstrip lines. 
These two signals are combined and processed by the combiner and envelope detector to generate a series of RF holograms, which are then transmitted to the control board for channel reconstruction and estimation. 
The channel reconstruction process is detailed in Fig. \ref{figExpSet}(d). Three holograms with distinct phase shifts are required for channel reconstruction. Prior to each hologram acquisition, the reference wave is phase-shifted by $\pi/2$, and the resulting hologram remains stable throughout the sampling interval in the case of coherent interferometric superposition of the signals.

To validate the accuracy of channel estimation, we estimate the direction of arrival (DOA) from the CSI acquired by the HIS using the Bartlett method, and then compare the estimated DOAs with the true values. 
The DOA estimation within the range of $-60^\circ$ to $60^\circ$ is evaluated in the experiment, and the results are presented in Fig. \ref{figExpSet}(e)-(g) and Fig. \ref{figMeas}. A comparison between the ideal and estimated results is shown in Fig. \ref{figExpSet}(e), demonstrating that the HIS achieves accurate angle-of-arrival determination. Fig. \ref{figExpSet}(f)-(g) illustrate the Bartlett estimation process for DOA of $40^\circ$, and $60^\circ$, respectively. These results confirm that the proposed interference-based channel sensing architecture is capable of high-precision channel estimation and shows strong potential for practical applications. Fig. \ref{figMeas} present the channel reconstruction results at different arrival angles, including the recovered signal phase and the three corresponding holograms. 
Although the experiments mentioned above were performed in a microwave chamber with near line-of-sight propagation, the HIS maintains accurate channel estimation capabilities in multipath environments.

\section{Discussion} \label{secDisc}
In this paper, we designed and experimentally demonstrated the holographic channel sensing architecture with HIS. We fabricated a 32-unit array equipped with multiple power detection sensors to enable RF chain-independent channel estimation. Using denoised holograms, we estimate the DOA via the Bartlett estimation method. Experimental results demonstrate the capability for accurate channel estimation and its potential for enabling cost-effective communication systems. Furthermore, we implemented and evaluated a prototype system using the proposed HIS platform, confirming its channel estimation performance. The proposed scheme offers several advantages, including simplified hardware architecture, reduced power consumption, and improved cost-effectiveness. Our work presents a novel approach to signal reception and channel estimation by processing only the power of received signals. Theoretical analysis and simulation results indicate that the proposed HIS-based communication architecture provides a simplified and intuitive framework for signal reception and channel estimation. Holographic interference surfaces overcome traditional hardware limitations on the number of antennas, offering a practical pathway for deploying massive MIMO arrays.

\section{Methods} \label{secMeth}

\subsection{Antenna radiation parameters}
The center frequency of the antenna unit is defined as the frequency at which the S11-parameter reaches its minimum value. The antenna bandwidth corresponds to the frequency range over which the S-parameter remains below -10 dB. The half-power beamwidth (HPBW) is defined as the angular width of the region where the gain exceeds -3 dB on the 2D radiation pattern. The cross-polarization discrimination is determined as the difference between the peak values of the co-polarization and cross-polarization curves on the 2D radiation pattern.

\subsection{Experimental verification}
The experiments are conducted in a microwave chamber to evaluate the channel estimation performance of the prototype, as shown in Fig. \ref{figExpSet}. The HIS was mounted on a rotation platform with its control board and power supply system. A standard gain horn antenna WR340 from A-INFO company was used to emit the object wave via an RF signal generator (Tektronix TSG 4106A).

\section{Data availability}
The authors declare that all relevant data are available in the paper and its Supplementary Information files, or from the corresponding author on request.

\section{Code availability}
The custom computer codes used in this study are available from the corresponding authors on request.

\bibliography{sn-bibliography}

\section{Acknowledgements}
This work was supported by the Fundamental Research Funds for the Central Universities and the National Natural Science Foundation of China under Grant 62071191.

\section{Author contributions}
H.Y. suggested the designs, planned and supervised the work. J.H. conceived the idea, carried out the analytical modeling and numerical simulations. R.Z. and J.W. fabricated the prototype, performed the theoretical analysis, and carried out the experimental measurements. J.H. and R.Z. performed the data analysis. H.Y., J.H. and R.Z. wrote the manuscript. All authors discussed the theoretical aspects and numerical simulations, interpreted the results, and reviewed the manuscript.

\end{document}